\documentclass[emulateapj,epsfig]{article}
\usepackage[onecolumn]{emulateapj}
\usepackage[]{epsfig}
\usepackage{graphicx}

\hoffset -1.5truecm \lefthead{Menci et al.}
\righthead{}
\def\newline{\hfil\break}
\begin{document}
\title{The Blast Wave Model for AGN Feedback: Effects on AGN Obscuration}
\author{N. Menci$^1$, F. Fiore$^1$, S. Puccetti$^{1,2}$, A. Cavaliere$^3$}
\affil{$^1$INAF - Osservatorio Astronomico di Roma, via di Frascati
33, I-00040 Monteporzio, Italy}
\affil{$^2$ ASI SDC, c/o ESRIN via Galileo Galilei, 00044 Frascati, Italy}
\affil{$^3$ Dip. Fisica,
Universita'di Roma Tor Vergata, via Ricerca Scientifica 1, 00133, Roma, Italy}

\smallskip

\begin{abstract}
We compute the effect of the galactic absorption on AGN emission in a
cosmological context by including a physical model for AGN feeding and feedback
in a semi-analytic model of galaxy formation. This is based on galaxy
interactions as triggers for AGN accretion, and on expanding blast waves as a
mechanism to propagate outwards the AGN energy injected into the interstellar
medium at the center of galaxies.

We first test our model against the observed number density of AGNs with
different intrinsic luminosity as a function of redshift. The model
yields a ''downsizing'' behavior in close agreement with
the observed one for $z\lesssim  2$. At higher redshifts, the model
predicts an overall abundance of AGNs (including Compton-thick sources)
larger than the observed Compton-thin sources by a factor $\approx 2$ for
$z\gtrsim 2$ and $L_X\leq 10^{44}$ erg/s. Thus, we expect that at such
luminosities and redshifts about $1/2$ of the total AGN population is contributed
by Compton-thick sources.

We then investigate the dependence of the absorbing column density $N_H$
associated to cold galactic gas (and responsible for the Compton-thin component
of the overall obscuration) on the AGN luminosity and
redshift. We find that the absorbed fraction of AGNs with $N_H\geq 10^{22}$
cm$^{-2}$ decreases with luminosity for  $z\leq 1$; in addition, the total
(integrated over luminosity) absorbed fraction increases with redshift up to
$z\approx 2$, and saturates to the value $\approx 0.8$ at higher redshifts. Finally, we predict the luminosity dependence of the absorbed fraction of AGNs
with $L_X\leq 3\,10^{44}$ erg/s to weaken with increasing redshift.

We compare our results with recent observations,
and discuss their implications in the context of cosmological models
of galaxy formation.

\end{abstract}

\keywords{galaxies: active --- galaxies: formation --- galaxies: evolution}

\section{Introduction}

The accretion which built up the supermassive black holes (SMBHs) now
hosted in many local galaxies is widely thought to be associated with a
sequence of output episodes observed as Active Galactic Nuclei
(AGNs). Since only a minority ($\sim 10^{-2}$, see Richstone et al.  1998)
of local galaxies host a currently active AGN, the corresponding  lifetimes
are estimated to be close to $\tau\sim 10^8$ yrs. Several
observations indicate the accretion episodes to be fundamentally
related to the galaxy growth; one such indication is provided by the
narrow scatter in the observed correlation of the SMBH mass $M_{BH}$
with its host galaxy mass $M$, when the former is in the range
$10^7\leq M_{BH}/M_{\odot}\leq 10^9$ (Ferrarese \& Merritt 2000,
Gebhardt et al. 2000). The emissions of AGNs thus may conceivably
constitute a probe for the history of accretion and growth of SMBHs,
and for its interplay with the galaxy building process.

The co-evolution of galaxies and AGNs and their so called
``downsizing'' (faster evolution for more luminous objects)
depends also on feedback between nuclear and other galactic
activities. In fact, the density of the high luminosity QSOs is peaked at high
redshift and declines strongly toward us; similarly, massive galaxies
are characterized by a star formation history peaked at high
redshifts. Luminous AGNs are efficient in "sterilizing" their massive
host galaxies by heating the interstellar matter through winds, shocks,
and high energy radiation, see Granato et al. (2004);  
Murray, Quataert, Thompson (2005); Hopkins et al. (2006); 
Bower et al. (2006); Menci et al. (2006). Intriguingly,
the latter authors found
that the bimodal color distribution of galaxies at z$\gtrsim 1.5$ can only
be explained if AGN feedback is considered.  In this picture an AGN
phase precedes the phase when a galaxy is caught in a passive state
with red optical-UV colors, most of the star-formation having been
inhibited by the AGN activity. Indeed, Pozzi et al. (2007) using
Spitzer photometry found that that a sample of optically obscured QSO
at z=1--2 are mainly hosted by red passive galaxies, suggesting a later stage
in their evolution.

On the other hand, at low redshift many weak AGNs have
been found in star-forming galaxies (Salim et al. 2007). In these
cases feedback from less powerful AGNs (the so-called ''radio mode'') 
is probably acting to
self-regulate accretion and star-formation, and cold gas is left
available for both processes for a much longer time (Croton et
al. 2006). The same cold gas can intercept the line of sight to the nucleus. 
Indeed, Compton-thin absorbers (with column densities $N_H\leq 10^{24}$ cm$^{-2}$)
may well be located in the galactic disk (Malkan, Gorjian, Tam 1998; Matt 2000; 
see also Ballantyne, Everett, Murray 2006). 
Therefore a natural expectation in this scenario is the fraction of obscured
AGNs to be large at low AGN luminosities. It is well known since the
pioneering work done with the {\it Einstein} satellite (Lawrence \&
Elvis 1982) and with optically and radio-selected AGNs (Lawrence 1991) 
that this fraction strongly decreases with increasing AGN
luminosity (see Ueda et al. 2003; La Franca et al. 2005; Gilli, Comastri, Hasinger
2007; Triester, Krolik, Dullemond 2008; Hasinger 2008).  A widely
shared view holds that the luminosity dependence of the obscured
fraction is related to the energy fed back by the AGNs onto the
surrounding gas that constitutes the interstellar medium (ISM). Given that
the AGN emission is proportional to the fraction of such a gas
available for accretion, a \textit{positive} correlation between
luminosity and absorption would be expected instead in the absence of
an energy feedback depleting the ISM after the onset of the AGN activity.

A mounting body of observations cogently indicates that strong nuclear feedback
is present in galaxies hosting AGNs (see for a review Elvis 2006 and
references therein).  On small (sub-pc) scales, the
observed X-ray absorption lines indicate the presence of outward winds
with velocities up to some $ 10^4$ km/s (Weymann 1981; 
Turnshek et al. 1988; Creenshaw et al. 2003; 
 Chartas et al. 2002; Pounds
et al. 2003, 2006; Risaliti et al. 2005b).  These likely 
originate from the acceleration of disk outflows due to the AGN
radiation field (Proga 2007 and references therein). On
larger scales, broad absorption lines in about 10\% of optically
luminous QSOs indicate fast outflows (up to 30,000 km/s).  Massive
(10-50 $M_{\odot}/yr$) flows of neutral gas with speed $\sim1000$ km/s
are observed through 21-cm absorption of radio-loud AGNs (see
Morganti, Tadhunter, Oosterloo 2005), indicating that AGNs have a
major effect on the circumnuclear gas in the central kiloparsec region
around  AGNs. On even larger scales of some $10^{2}$ kpc, the presence of
AGN-induced outflows is revealed by X-ray observations of the
intra-cluster medium (see McNamara \& Nulsen 2007 for a review)
showing cavities and expanding shocks with Mach numbers ranging from
$\approx 1.5$ to $\sim 8$. How the outflows produced in the innermost
regions of the active galaxies are transported outwards to affect such
large scales is still matter of investigation; buoyant bubbles (see
Reynolds, Heinz \& Begelman 2001, Churazov 2001) and expanding blast
waves (see Cavaliere, Lapi \& Menci 2002; Lapi, Cavaliere \& Menci 2005)
constitute viable mechanisms for such a transport.

Nuclear obscuration is directly linked to AGN feedback, since the same
gas (and dust) which feed the AGN output may well be
responsible for its obscuration. Therefore, modelling AGN obscuration
is essential to connect the observed AGN properties to the
accretion history of SMBHs over cosmological time. This is a significant
theoretical challenge as it requires not only connecting the AGN
evolution to the galaxy formation and growth, but also implementing in
a model a self-consistent description of the AGN feedback on the
galactic gas. Indeed, very few attempts have been
carried out in this direction so far. 
An attempt to include AGN absorption into an {\it ab initio}
galaxy formation model has been proposed by Nulsen \& Fabian
(2000), by relating both the SMBH fueling and the AGN
absorption to the cooling flows associated with the hot gas pervading
the growing dark matter haloes, but this assumption did not lead to a 
full description of the statistical distribution of QSO luminosity.

Here we develop our semi-analytic model of hierachical galaxy
formation and AGN evolution (see Menci 2006) to self-consistently
include the absorption of AGNs, with the aim of investigating the
cosmic evolution of the latter and its dependence on AGN
properties like luminosity and redshifts.
The model is suited to our scope as it includes a detailed 
treatment of the feedback on the interstellar gas. Our aim is
to investigate the dependence of AGN {\it absorption} on luminosity $L$ and
redshift $z$ arising in hierarchical galaxy formation scenarios that
include an effective description of the evolution of AGNs and of their
feedback.

The plan of the paper is as follows: in Sect. 2 we describe our model
for galaxy formation and the associated AGN evolution. Sect. 3 is
focussed on describing our treatment of the AGN feedback onto the
interstellar gas and  its effects on the AGN absorption. In Sect. 4
we test our feedback-inclusive model for AGN evolution by comparing
its outcomes with the redshift distribution of the number density of
AGNs for different X-ray luminosities. Our results on the luminosity
and redshift dependence of the absorbed fraction of AGNs are shown and
discussed in Sect. 5. Sect. 6 is devoted to summarize our conclusions.

\section{The Model}

The semi-analytic model we develop and use  connects, within a cosmological framework,
the accretion onto SMBHs and the ensuing AGN activities with the  evolution of galaxies.

\subsection{Hierarchical Galaxy Evolution}

Galaxy formation and evolution is driven by the collapse and growth of dark
matter (DM) haloes, which originate by gravitational instability of  overdense
regions in the primordial DM density field. This is taken to be a random,
Gaussian  density field with Cold Dark Matter (CDM) power spectrum within the
''concordance cosmology" (Spergel et al. 2006) for which we adopt round
parameters  $\Omega_{\Lambda}=0.7$, $\Omega_{0}=0.3$, baryonic density
$\Omega_b=0.04$ and Hubble constant (in units of 100 km/s/Mpc) $h=0.7$. The
normalization of the spectrum is taken to be $\sigma_8=0.9$ in terms of the variance
of the field smoothed over regions of 8 $h^{-1}$ Mpc.

As  cosmic time increases, larger and larger regions of the density field
collapse, and  eventually lead to the formation of groups and clusters of
galaxies; previously formed, galactic size  condensations are enclosed. In
closer detail , the process implies  not only smooth mass inflow, but also
merging and coalescence of smaller condensations. The corresponding merging rates
of the DM haloes are provided by the Extended Press \& Schechter formalism (see
Bond et al. 1991; Lacey \& Cole 1993).  The clumps included into larger DM haloes
may survive as satellites, or merge to form larger galaxies due to binary
aggregations,  or coalesce into the central dominant galaxy due to dynamical
friction; these processes take place over timescales that grow longer over
cosmic time, so the number of satellite galaxies increases as the DM host haloes
grow from groups to clusters. All the above processes are implemented in our
model  as described in detail in Menci et al. (2005, 2006), based on canonical
prescriptions of semianalytic modeling.

The radiative gas cooling, the ensuing star formation and the
Supernova events with the associated feedback occurring  in the growing
DM haloes (with mass $m$ and circular velocity $v$) are described in
our previous papers (e.g., Menci et al. 2005). The cooled gas with mass
$m_c$ settles into a rotationally supported disk with radius $r_d$
(typically ranging from $1$ to $5 $ kpc), rotation velocity $v_d$
and dynamical time $t_d=r_d/v_d$. The gas gradually condenses  into
stars at a rate $\dot m_*\propto m_c/t_d$; the stellar ensuing feedback
returns part of the cooled gas to the hot gas phase
with mass $m_h$ at the virial temperature of the halo. An additional
channel for star formation implemented in the model is provided by
interaction-driven starbursts, triggered not only by merging but
also by fly-by events between galaxies; such a star formation mode
provides an important contribution to the early formation of stars
in massive galaxies, as described in detail in Menci et al. (2004,
2005).

\subsection{Accretion onto SMBHs and AGN emission}

The model also includes a treatment of SMBHs growing at the centre of galaxies
by interaction-triggered inflow of cold gas, following the physical model proposed
by Cavaliere \& Vittorini (2000) and implemented in Menci et al. (2006). The
accretion of cold gas is triggered by galaxy encounters (both of fly-by and
of merging kind), which destabilizes part of the cold gas available by inducing
loss af angular momentum. In fact, small scale (0.1-a few $kpc$) regions are likely to have 
disk geometry if they are to efficiently remove angular momentum and convey to 
the (unresolved) pc scales the gas provided on larger scales (these 
may be isotropized by head-on, major merging events). 

The rate of such  interactions is given by Menci et
al. (2003) in the form
\begin{equation}
\tau_r^{-1}=n_T\,\Sigma (r_t,v,V_{rel})\,V_{rel}.
\end{equation}
Here $n_T$ is the number density of galaxies in the same halo and
$V_{rel}$ is their relative velocity. Encounters effective for
angular momentum transfer require i) the interaction time to be
comparable with the internal dynamical times of the partner galaxies
(a resonance condition), ii) the orbital specific energy of the
partners not to exceed the sum of their specific internal
gravitational energies. The cross section $\Sigma$ for such
encounters is given by Saslaw (1985) in terms of the tidal radius
$r_t$ associated to a galaxy with given circular velocity $v$ (see
Menci et al. 2003, 2004).

The fraction of cold gas accreted by the BH in an interaction event is computed
in terms the  variation $\Delta j$ of the specific angular momentum $j\approx
Gm/v_d$ of the gas to read (Menci et al. 2003)
\begin{equation}
f_{acc}\approx 10^{-1}\,
\Big|{\Delta j\over j}\Big|=
10^{-1}\Big\langle {m'\over m}\,{r_d\over b}\,{v_d\over V_{rel}}\Big\rangle\, .
\end{equation}
Here $b$ is the impact parameter, evaluated as the average distance of the
galaxies in the halo. Also, $m'$ is the mass of the  partner galaxy in the
interaction,  and the average runs over the probability of finding such a galaxy
in the same halo where the galaxy with mass $m$ is located.
The values of the quantities involved in the average are computed from the 
semi-analytic model recalled in Sect. 2.1, and yield values of $f_{acc}\lesssim 10^{-2}$. 
For minor merging events and for the encounters among galaxies with very unequal mass ratios
$m'\ll m$, dominating the statistics in all hierarchical models of galaxy formation, the 
accreted fraction takes values $10^{-3}\lesssim f_{acc}\lesssim 10^{-2}$. 

The average amount of cold gas accreted during an accretion episode is thus
$\Delta m_{acc}=f_{acc}\,m_c$, and the duration of an accretion episode, i.e.,
the timescale for the QSO or AGN to shine, is assumed to be the crossing time
$\tau=r_d/v_d$ for the destabilized cold gas component.

The  time-averaged bolometric luminosity so produced by a QSO hosted in a given galaxy 
is then provided  by
\begin{equation}
L={\eta\,c^2\Delta m_{acc}\over \tau} ~.
\end{equation}
We adopt an energy-conversion efficiency $\eta= 0.1$ (see Yu \&
Tremaine 2002), and derive the X-ray luminosities $L_X$ in the 2-10
keV band  from the bolometric corrections given in Marconi et al.
(2004). The SMBH mass $m_{BH}$ grows mainly through accretion
episodes as described above, besides  coalescence with other SMBHs
during galaxy merging. As initial condition, we assume small seed
BHs of mass $10^2\,M_{\odot}$ (Madau \& Rees 2001) to be initially
present in all galaxy progenitors; our results are insensitive to
the specific value as long as it is smaller than some
$10^5\,M_{\odot}$.

In our Monte Carlo model, at each time step we assign to a  galaxy
the interaction probability corresponding to the rate given in eq. (1).
According to it,  we assign to the galaxy an active SMBH accretion event of
duration $\tau$.  Then we compute the accreted cold gas and the
associated AGN emission through equations (2) and (3).

\section{The AGN feedback and the column density of absorbing gas}

The model for AGN evolution presented here -- relative  to previous
implementation by Menci et al. (2003-2004) -- is complemented with
the feedback from AGNs onto the surrounding ISM. This affects: a)
the available cold gas left over  in the galactic disk after each
AGN event, that determines the subsequent gas accretion history; b)
the density distribution of the galactic gas during each AGN event,
that determines  the AGN absorption.

\subsection[]{The Blast Wave Model for the AGN Feedback}
As mentioned in the Introduction, fast winds with velocity up to
$10^{-1}c$ are observed in the central regions of AGNs; they likely
originate from the acceleration of disk outflows by the AGN
radiation field (see Begelman 2003 for a review), and affect the
environment in the host galaxy and beyond, leaving imprints out to
large scales of some $10^2$ kpc  in the intra-cluster medium (ICM).

A detailed model for the transport of energy from the inner, outflow
region to the large scales has been developed by Cavaliere, Lapi \&
Menci (2002), and Lapi, Cavaliere  \& Menci (2005). Central, highly
supersonic outflows compress the gas into a blast wave terminated by
a leading shock front, which  moves outwards with a lower but still
supersonic speed and sweeps  out the surrounding medium. Eventually,
this is expelled from the galaxy;  in the case of powerful shocks,
it is expelled  even  from a  group or poor cluster hosting the galaxy.

The key quantity determining all  shock properties is the total energy $\Delta E$
injected by AGNs into the surrounding gas. This is computed as
\begin{equation}
\Delta E = \epsilon_{AGN}\,\eta\,c^2\,\Delta m_{acc}
\end{equation}
(see Sect. 2.2) for each SMBH accretion episode in our Monte Carlo
simulation; the value of the energy feedback efficiency for coupling with
the surrounding gas is taken  as
$\epsilon_{AGN}=5\, 10^{-2}$, consistent with the values required to
match the X-ray properties of the ICM  in clusters of galaxies (see
Cavaliere, Lapi \& Menci 2002). This is also consistent with the
observations of wind speeds up to $v_w\approx 0.1\,c$ in the central
regions, that yield $\epsilon_{AGN}\approx v_w/2c\approx 0.05$ by
momentum conservation between photons and particles
(see Chartas 2002, Pounds 2003); this value has been also adopted
in a number of simulations (e.g., Di
Matteo, Springel  \& Hernquist 2005) and semi-analytic models
 of galaxy formation (e.g., Menci et al. 2006).

The blast expands  into the ISM or the ICM  as  described by the
hydrodynamical equations; these are to include the effects  not
only   of  initial density gradient, but also those of  upstream
pressure and DM gravity,  clearly important quantities in galaxies,
as discussed by Lapi, Cavaliere, Menci (2005);
the solutions show in detail how the
perturbed gas is confined to an expanding  shell bounded by an outer
shock at the  radius $R_s(t)$ which sweeps out the gas surrounding
the AGN. The overall effects have been emulated in the simulations
by Di Matteo et al. (2005), performed for specific  cases of major
mergers. Our treatment covers also the less energetic but more
frequent events, with   accretion rates and AGN energy inputs
$\Delta E$ provided by eqs. (1) to (4).  On the other hand, 
our model assumes spherical symmetry; 
note however that this approximatively holds in the central 
regions (the ones mainly affecting our results, as we shall show below) 
due the exponential decline of the gas density. 
In addition, both 
the effects discussed in points 1) and 2) of the Appendix (see below Sect. 3.2) 
make the average absorption even less dependent on the disk outside the central region.
As for the outflow geometry,  our model also applies to segments of spherical shells expanding outwards, originated by winds driven by a spherically symmetric AGN radiation field; 
these segments may not cover uniformly the central source, as in the conical distribution of Elvis et al. (2000). While its is true that random line-of-sights may not intercept some of these segments, the random orientation of the internal BH accretion disk with respect to the outer, galactic disk (Gallimore,  Baum,  O'Dea 1997; Nagar \& Wilson 1999; Thean et al. 2001) makes our results realistic on average. Finally, we do not consider clumpy outflows; however, these are not expected to affect the dependence of the column density distribution on the AGN luminosity and redshift, though they could broaden the column density distributions that we derive from our model.
 
Lapi et al. (2005) showed in detail how  the hydrodynamical
equations for the finite amplitude gas perturbation constituting the
blast wave, supplemented with the Rankine-Hugoniot boundary
conditions at the shock, still admit  self-similar solutions (see
Sedov 1959) when  gravity,  initial gradients and pressure are
included. These solutions at the distance $r$ from the center
are expressed in terms of $r/R_s$, where
$R_s(t)$ is the shock position after a time
$t$ from an  AGN outburst; this is given by the eq. (C7) in Lapi et al.
(2005), that we recast in the form
\begin{equation}
R_s(t)=v_d\,t_d\,\Big[{5\,\pi \omega^2 \over 24 \pi (\omega-1)}\Big]^{1/\omega}
{\mathcal M}^{2/\omega}\Big[{t\over t_d}\Big]^{2/\omega},
\end{equation}
where ${\mathcal M}$ is the Mach number ${\mathcal M}=v/c_s(R_s(t))$.

Here the initial gas density has a radial
profile given by  a power law $\rho\propto r^{\omega}$  with the exponent in the
realistic  range $2\leq \omega <2.5$. While self-similar, analytical solutions
for the shock expansion law can only be found  for spherical (or in general 1D)
symmetry and power law gradients, we shall apply eq. (5) also to blasts
expanding in the inner regions of a galactic disk, where the actual  density
profile follows an exponential law $\rho=(1/2\,h)\,\sigma_o\,exp(-r/r_d)$ ($h$
is the disk thickness and $\sigma_0$ is the central surface density), with a
cutoff in the vertical direction corresponding to the thickness of the disk. To
use the above self-similar blast we construct  piecewise power-law approximations of
the above exponential density by subdividing the radial disk coordinate into a
sequence of shells; within each shell the exponential decline is approximated
with a power-law with a different exponent $\omega$.  The expansion of the blast
in each shell will be computed on using eq. (5) with the appropriate value of
$\omega$.
 
Self-similarity imply a definite time behavior for the AGN energy injection
which is determined by the exponent $\omega$ defining the density decline, normalized  
to reach its total value $\Delta E$ by the end of the accretion episode. 
In turn, this implies a well-defined form for the time decay of the AGN luminosity, 
as shown in the Appendix. However, in the self-similar solutions the ratio 
$\Delta E/E$ between the energy injected up to a time $t$ and the energy of the ISM 
within the shock radius $R_s(t)$ is shown to be independent of time and position
(see Appendix), and thus constitutes the basic quantity marking the strength 
of the shock. In fact. 
 the Mach number is shown to be simply related to $\Delta E/E$ which we compute 
at the final time of the AGN active phase. 
In the following,  we shall take ${\mathcal M}^2=1+\Delta
E/E$ as derived by  Lapi et al. (2005).

In terms of the rescaled variable $r/R_s$, the previous authors  derived the
density distribution;  this is found to be confined to  a shell close to $R_s$
between the outer shock and an inner "piston"  at a position $\lambda\,R_s(t)$
($\lambda < 0.7$) with a width  $(1-\lambda)\,R_s$ weakly dependent on the Mach
number. Inside  the shell little  gas is  left over  to absorb the AGN
radiation.

\subsection[]{The Effect of a Blastwave on the Column Density of the Absorbing Galactic Gas}

The amount of unshocked gas absorbing the AGN emission will be given, for a
given line-of sight, by the time-dependent fraction still unperturbed outside
the shock located at  $r= R_s(t)$ (see fig. 1). The faster the shock, the lower
will be the fraction of still unperturbed gas outside $R_s(t)$ after a time
$t$ from an  AGN outburst.

The dependence of $R_s(t)$ in eq. (5) on the AGN luminosity will then affect the
absorption by  the unperturbed gas;  in particular, for the  case $\omega=2$
(corresponding to the standard isothermal sphere) the relation $R_s(t)\propto
{\mathcal M}\,t\propto (\Delta E/E)^{1/2}\,t$ holds, yielding {\it lower} absorbing
gas fraction outside $R_s(t)$ for {\it brighter} AGNs at a fixed time $t$. 

An additional contribution to the absorption will be given by the gas compressed
by the shock in  the  shell of   width $(1-\lambda)\,R_s$ (see  Lapi, Cavaliere
\& Menci 2005), where $\lambda$ can  be also expressed in terms of the relevant
ratio $\Delta E/E$; however, we shall show  this second contribution to  depend
only weakly  on the AGN luminosity.

\begin{center}
\vspace{-0.cm}
\scalebox{0.73}[0.73]{{\includegraphics{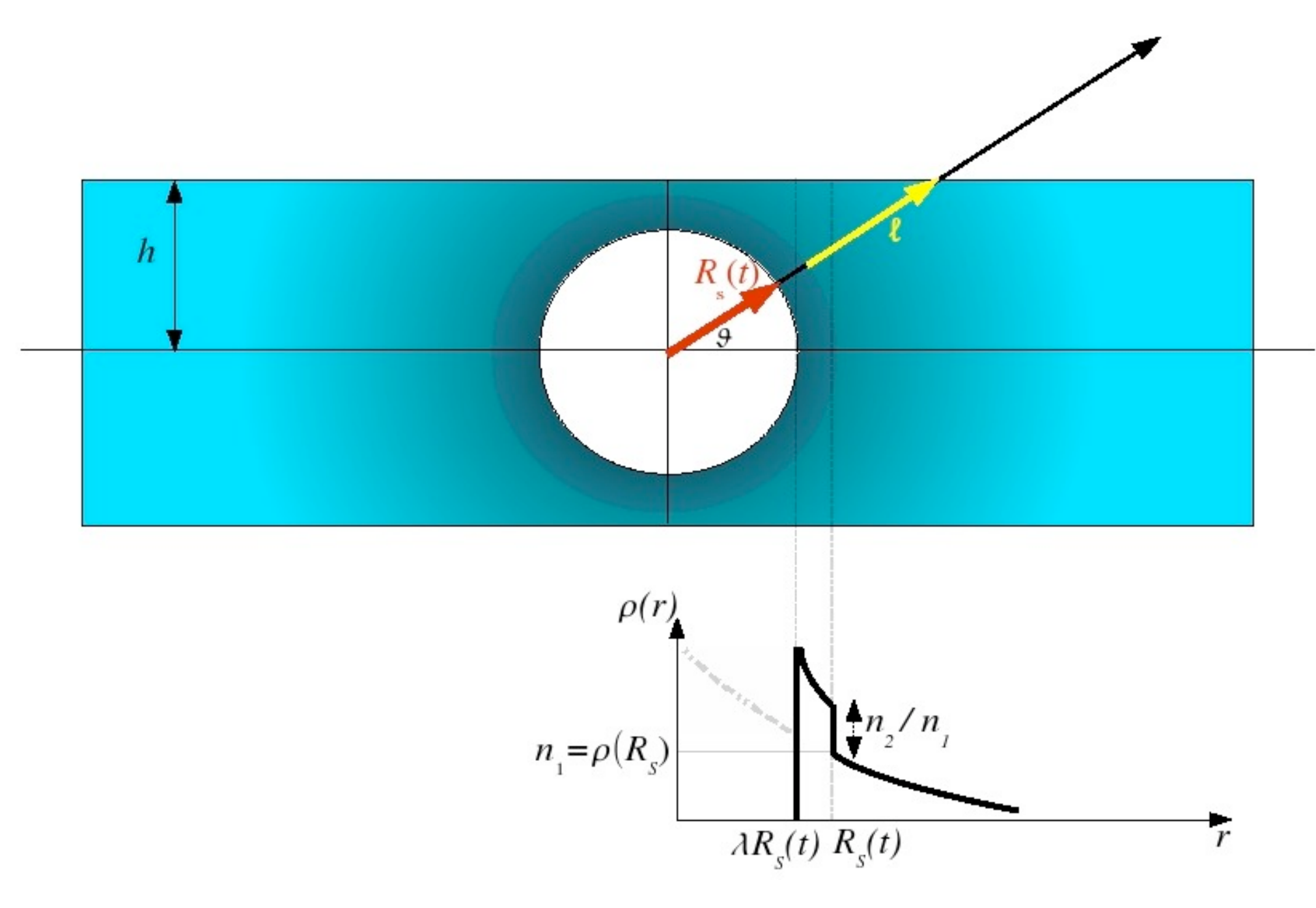}}}
\end{center} {\footnotesize \vspace{-0.4cm }
Fig. 1. - A schematic representation of the
effect of the blast wave induce by AGN feedback on the density
distribution of the interstellar gas. The shock radius $R_s(t)$
expands outwards, compressing the swept gas into a thin shell
(represented in darker colour) with width $R_s(1-\lambda)$, and
leaving a cavity inside. Such a density distribution $\rho(r)$ is
also plotted at the bottom. A line-of-sight (at an
angle $\theta$ with respect to the plane of the galactic disk) is
also represented in the scheme, along with the length $\ell$ of the
line of sight intercepting the unperturbed gas outward of the shock
front.
 \vspace{0.4cm}}

We implement the above model for AGN feedback and for absorption in
our SAM as follows. \newline \newline 0) For each galaxy in our
Monte Carlo simulations, with disk exponential scale length $r_d$ and
total disk mass $m_c$ computed as described in \S  2.1, we assume
for the density distribution of the  unperturbed gas the simplified form
$\rho=\rho_o\,exp(-r/r_d)$ (where $r$ is the distance from the
centre of the galaxy) with a cutoff in the vertical direction
(perpendicular to the disk) at $r=h$ corresponding to the disk
thickness. \newline The value of the vertical scale height $h$ for
the {\it gaseous} disk is taken to be $r_d/15$ (see Narayan \& Jog
2002),  corresponding to $\sim 200$ pc for a typical $L_*$ galaxy.
With such a low value of $h$, our assumed density for the gaseous
disk closely follows a constant vertical density distribution with a
radial exponential profile $exp(-x/r_d)$, $x$ being the distance
from the centre in the plane of the disk; the deviations are  $\leq
14 \% $ in the central region,  and $\leq 1 \%$ at $x=r_d$. These
are  both smaller than the present uncertainty on the vertical
distribution of the density of the gaseous disks; on the other hand,
the above schematic  form for the gas density distribution allows us
to apply a simple  anaytical description for  the evolution of the
blast. The density distribution is normalized as to recover the
total gas mass $m_c$ when integrated over the disk volume.
 \newline
 \newline
1) We construct a piecewise power-law approximation of the above
exponential density distribution, by subdividing the radial
coordinate into shells; in each shell we approximate the exponential
shape  $exp(-r/r_d)$ with a power-law $A\,q^{\omega}\,s^{-\omega}$
where  $s=r+q$, where $q$, $A$, and $\omega$  are parameters that we
adjust to optimize the fit in each radial shell. We find that, on
retaining  $q=1.8\,r_d$ at all radii,  the following set of radial
shells and  parameters provides a good fit to the exponential law:
\newline $A=1$, $\omega=2$ for $r_m\leq r\leq r_d/2$,
\newline $A=1.23$, $\omega=2.3$ for $r_d/2\leq r\leq 3\,r_d/2$,
\newline $A=1$, $\omega=2.5$ for
$r\geq 3\,r_d/2$; \newline
here the minimal  radius considered is
$r_m=50$ pc. With the above choice, the r.m.s. deviation in the fits
are $\Delta\leq 3\%$ within the inner region ($r\leq 10 r_d$, the
dominant  range for the AGN abosrption),   and $\Delta\leq 30\%$ for
$r\geq 10 r_d$ (a region irrelevant to our computation of
the AGN absorbtion). The temperature of the gas in the disk is
assumed to be $T_d=10^4$ K.
\newline
 \newline
2) When a  SMBH starts its accretion phase triggered by a close
interaction of the host galaxy with a  companion, we compute the
corresponding AGN total energy injection $\Delta E$. For the sake of 
simplicity, we shall label each AGN with a constant luminosity $L$ after eq. (3), 
although a full consistency with the self-similar solutions adopted for the 
blastwave expansion would require a time-decay of the luminosity related to the 
exponent $\omega$. However, we show in the Appendix that assuming a constant $L$ 
yields results not distinguishable from those obtained by assigning the AGNs the exact, 
time-dependent luminosity.
\newline
\newline
3) During the time interval $\tau$ we follow the expansion of the shock given
by eq. (5) for each radial shell  defined under  1), on  using the appropriate
value of $\omega$ given by the fitting procedure described above. We compute the
amount of gas still unperturbed outwards of  the shock position $R_s(t)$ at a
time $t$ within the  interval $\tau$; the unshocked gas distribution is  given
by the density profile  $\rho(r)$ defined under  (1), with $r\geq R_s(t)$ (see
fig. 1).
\newline
\newline
4) We extract a random line-of-sight angle $\theta$ defining the disk
inclination to the observer; at a time $t$ within the interval $\tau$
corresponding to the active AGN,  we compute the column density corresponding to
the gas outside the shock position along the selected line-of-sight as
\begin{equation}
N_H=\int_{R_s(t)}^{h/sin{\theta}} \rho(r)\,d\ell ~.
\end{equation}
5) If for the chosen line-of-sight the shock position is within the
disk, we add the absorption column density corresponding to the gas
compressed by the shock (see fig. 1).  Lapi et al. (2005) showed
that the density of the gas swept by the blast  and  compressed to a
shell of width $(1-\lambda)\,R_s$ rises toward the center as a
function $D(r/R_s(t))$ diverging weakly (but still integrating to a
limited mass) at the piston position $\lambda\,R_s(t)$. On adopting
the behaviour
$D(r/R_s(t))=[(r/\lambda\,R_s)-1]^{(\omega-6)/3(7-\omega)}$
(strictly valid  on approaching the piston position), and
integrating over $r$ within the shocked shell, the corresponding
column density is
\begin{equation}
N_{H, shock}={15-5\omega\over 3(7-2\omega)}\,\Big[{1\over \lambda}-1\Big]^{15-5\omega\over 3(7-2\omega)}\,{n_2\over n_1}\,n_1\,R_s(t)~.
\end{equation}
Here $n_2/n_1$ is the density jump at the shock, and $n_1=\rho (R_s)$ is the
unperturbed gas density just outside the shock front. Both the density jump
$n_2/n_1$ and the $\lambda$ grow slowly  with increasing Mach number
$\mathcal{M}$ (in turn a function of $\Delta E/E$), as given in Lapi et al.
(2005); both  quantities saturate to constant values ($n_2/n_1=4$ and
$\lambda\approx 0.7$) in the limit of very strong shocks ($\mathcal{M}\gg 1$).

In sum, the blastwave has two effects: on the one hand, it depletes
the internal regions    thus decreasing the column density of the
absorber (as computed at point 4 above); on the other hand, it
increases the density inside a thin layer close to the piston (an
effect computed at point 5 above). The first effect however
dominates, since the gas initially distributed in the
denser central region is spread out by the blast into an expanding
spherical shell; thus, the integration along
the line of sight intercepts only a tiny fraction of the gas
initially concentrated in the core of the distribution. 
In addition, the minor contribution to the column 
density from eq. (7) represents an upper limit, since 
the gas temperature in the blast wave is
increased to values (given in eq. C8 of Lapi, Cavaliere, Menci 2005)
which generally exceed the ionization temperature of the gas and
thus limit the absorption.
At the end of the AGN accretion episode (i.e., at
$t=\tau$) only a fraction of the initial cold gas mass $m_c$ will be
left in the galaxy, as discussed in detail in Cavaliere et al.
(2002) and Lapi et al. (2005). Note that, the brighter is the AGN,
the larger is the injected energy  $\Delta E$, the faster is the
shock espansion $R_s(t)$ (see eq. 5), the lower will be the
resulting column density at a given time after the start of the
outburst (see eq. 6).

Thus the evolution of the gas distribution and hence of the gas absorption
during the AGN event is best  described in terms of the key ratio $\Delta E/E$
which determines all  the relevant  quantities of the blast: the expansion of
the shock front $R_s(t)$, the thickness of the shocked shell $(1-\lambda)$, and
the density jump at the shock. Such a ratio is computed in the model for each
galaxy when an active accretion episode is triggered by a close encounter.

The effects of the above detailed model for the AGN feedback on the evolution of
the cold galactic gas and on the galactic absorption of the AGN emission are
given  next.

\section{Results}

Here we present our results concerning the evolution of the AGN number density
and the obscuration by galactic interstellar gas.
We do not provide a description of absorption
due to the ''clouds'' within the central regions of AGNs, 
so we cannot disentangle Compton-thick sources (where the obscuration is
presumably provided by gas in regions $r<10$ $pc$) from the absorbed
population we obtain from our model. Indeed, 
the shock may originate at any point within our resolution scale (50 $pc$), 
and it may lie inside or outside the Compton-thick region. 
Therefore we cannot make any strong statement about the dependency of 
Compton-thick absorption on the outflow and hence on the AGN luminosity.

Before addressing the problem of the luminosity and redshift dependence of
AGN absorption, we test the basic predictions of our model with recent data.
To this aim, we first compare in fig. 2 the predicted evolution of the cosmic
density of AGN with different {\it intrinsic} luminosity $L$ with
data from La Franca et al. (2005). Note that the model reproduces the
marked {\it downsizing} of AGNs, i.e., the faster redshift evolution of brighter AGNs
compared to the fainter sources (a behaviour already obtained in  semi-analytic models, 
see Menci et al. 2004; Fontanot et al. 2006). In the model, such an
effect is related to the faster consumption of gas in the
progenitors of massive galaxies at high redshifts $z\gtrsim 3$; in
fact, these are formed in biased high regions of the cosmological 
density field, where the early cooling of galactic gas provides
early reservoirs for BH accretion, and the galaxy merging and
interactions are enhanced. When such progenitors assemble to form a
massive galaxy (and correspondingly a bright AGN when the BH is
actively accretiing) at $z\lesssim 2$, the available galactic
gas has largely been converted into stars or accreted onto the BH at
higher redshifts. Conversely, low-mass galaxies retain a large
fraction of their gas down to low redshifts, and this results into
prolonged star formation (yielding blue colors as observed) 
allowing for effective BH accretion down to $z\approx 0$.

\begin{center}
\vspace{0.2cm}
\scalebox{0.5}[0.5]{{\includegraphics{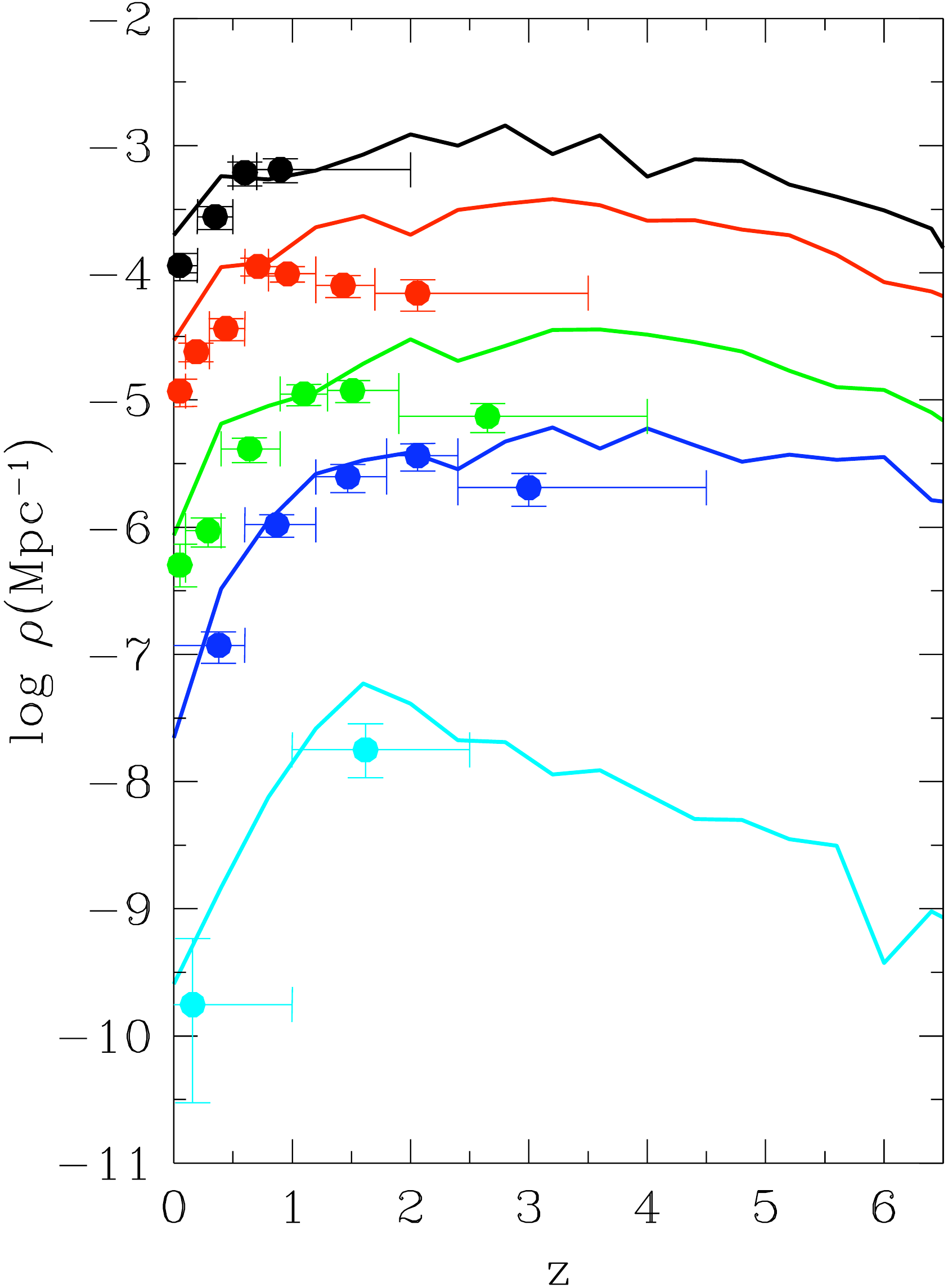}}}
\end{center} {\footnotesize \vspace{0.cm }
Fig. 2. - 
The evolution of the cosmic density of AGNs with X-ray luminosities
(in the $2-10$ keV band)
in 5 different luminosity bins: $42\leq$ log $L$/erg s$^{-1}< 43$ (top, blak line),
$43\leq $ log $L$/erg s$^{-1}< 44$ (red line), $44\leq$ log $L$/erg s$^{-1}< 44.5$ (green line),
$44.5\leq$ log $L$/erg s$^{-1}< 45$ (blue line), log $L$/erg s$^{-1}\geq 45$ (bottom cyan line).
Note that the the plotted absorbed fraction includes the Compton-thick sources.
 \vspace{0.4cm}}

Note that at $z\gtrsim 2$ the model predictions lie above the observed points by a factor $\approx 2$ for low luminosity AGNs with $L_X\leq 10^{43}$ erg/s. This, at least in part, is because
the model predictions in fig. 2 concern all AGNs, regardless of whether they are Compton-thin or Compton-thick,since we cannot disentangle the two populations in our model, while the data we compare with include only  Compton-thin sources.  
A complementary check would be constituted by the 
X-ray background at its 30 keV peak (where observations are not affected by Copton-thick obscuration). However this does not actually constitute a strong constraint for the model, since it is mainly contributed by objects at $z\leq 2$ (Gilli, Comastri, Hasinger 2007) whose observed density is well reproduced by the model. We have repetedly checked since Menci et al. (2003; 2004)  that  the basic features of our model are consistent with the constraint set by the observed X-ray background (see Barcons, Mateos, Ceballos, 2000) and that the predicted global history of BH assembly is consistent with the local observed SMBH mass function 
(Marconi et al. 2004). 

Thus the comparison between model predictions and observation strongly suggests the existence
of a relevant fraction of Compton-thick sources at $z\gtrsim 1.5$ (as found by Martinez-Sansigre et al. 2005, 2007;
Fiore et al. 2008a,b). Note however that a complementary possibility, which could partly account for the model
overestimate of low-luminosity sources at $z\gtrsim 2$, is that the BH growth in
small-mass galaxy halos (DM masses $M\leq 10^9\,M_{\odot}$) be inhibited by some
process not included in the model; we shall comment further this point in the Conclusions.

Finally, we stress that the model predicts at $z\gtrsim 5$ a density of very luminous-objects  
($L_X\geq 10^{45}$ erg/s) similar to that at $z\approx 1$. These are the X-ray counterparts of the
luminous $M_B\lesssim -27.2$ optical QSOs observed up to $z\approx 6.5$
(see Fan et al. 2004; Richards et al. 2006) and imply that fast 
building up of massive BH at early epochs is possible in hierarchical scenarios
(as long as a constant mass-energy conversion factor $\eta\approx 0.1$ holds, see
discussion in the final Section). In turn, this is a consequence of the speed up of
structure growth (enhanced merging rate) for haloes collapsed in biased, high-density
regions of the primordial density field, which constitutes a natural feature of hierarchical
scenarios.

We now turn to investigating the predictions of our model
concerning the luminosity and the redshift  dependence of the
absorption due to galactic gas. In fig. 3 we show
the predicted distribution of the column densities of the absorbers
for low redshift ($z\leq 1$) AGNs with different luminosity, and
compare it with data from La Franca et al. (2005), and with expectations based
on AGN synthesis models for the cosmic X-ray background (Gilli et al. 2007).
The rapid expansion of the blast wave produced by bright AGNs yields a low
fraction of absorbed ($N_H\geq 10^{23}$ cm$^{-2}$), bright ($L_X\geq
10^{45.5}$ erg/s) AGNs, while at lower luminosity the $N_H$
distribution shifts to larger values.
Note that the above relation between the luminosity and the obscuring column density
holds only after averaging over {\it time} after the start of the AGN active phase and
over the {\it orientations} of the galactic disk, the two fundamental
parameters that define the absorption properties of any single AGN in our model.
The probability for  a line-of-sight to intercept a given amount of
galactic gas is independent of the AGN luminosity (although models predicting such a dependence
have been proposed, see Lamastra, Perola \& Matt 2006), while the probability to observe an AGN
at a time when the blast wave has depleted the gas content of the galaxy
is larger for luminous AGNs, since these produce faster blast waves.

\begin{center}
\vspace{0.cm}
\scalebox{0.6}[0.6]{\rotatebox{-90}{\includegraphics{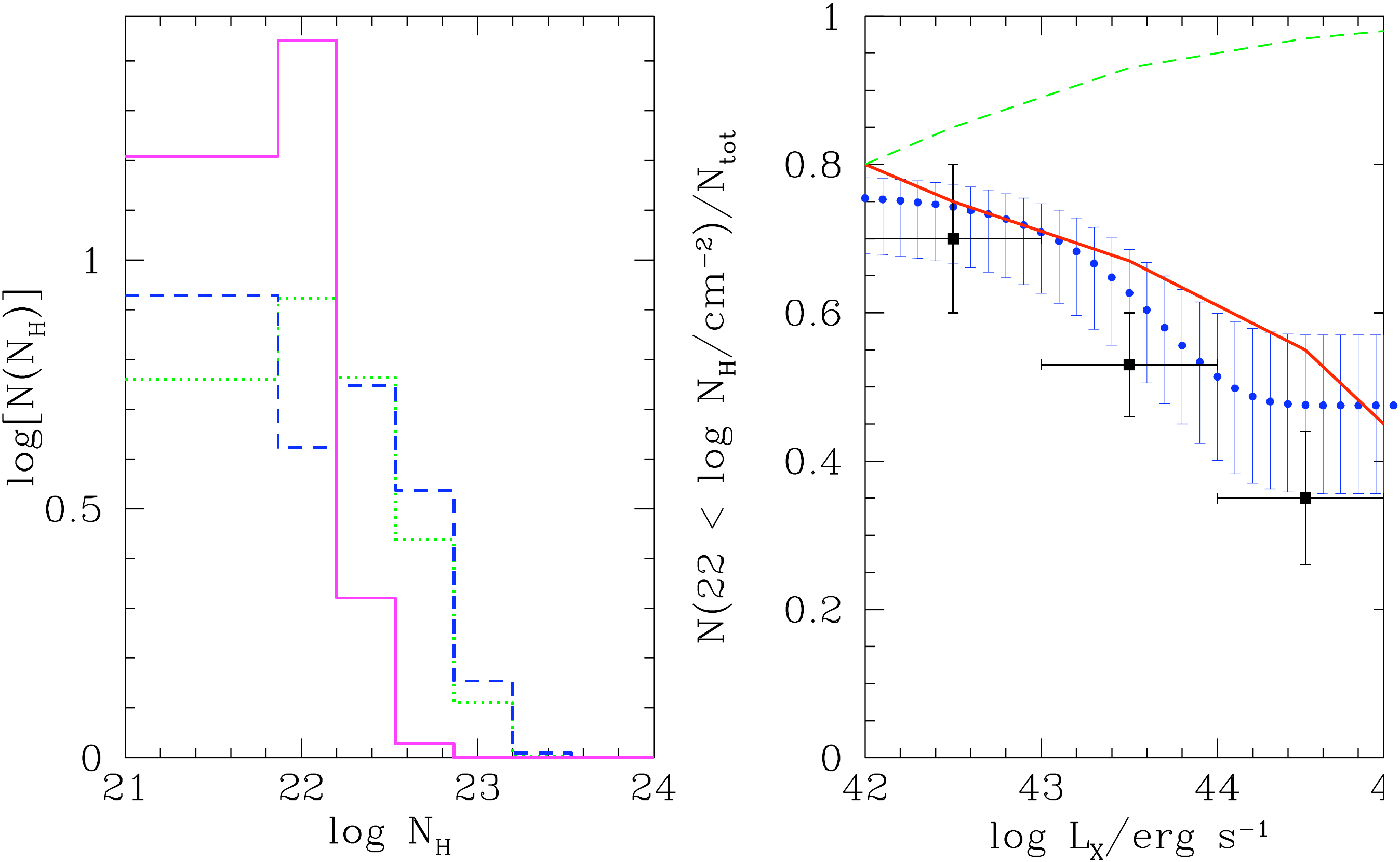}}}
\end{center} {\footnotesize \vspace{0.cm }
Fig. 3. - 
Left panel: The distribution of
column densities $N_H$ at $z\leq 1$ for AGN with different luminosities:
42$\leq$ log $L_X$/erg s$^{-1}\leq$ 43 (green dotted line), 43$\leq$
log $L_X$/erg s$^{-1}\leq$ 44 (red dashed line), log $L_X$/erg
s$^{-1}\geq$ 44 (violet solid line). The lowest $N_H$ bin
actually includes the contribution from all absorbing column
densities $N_H \leq 10^{22}$ cm$^{-2}$. \newline Right panel: The
predicted fraction of absorbed AGNs with $N_H>10^{22}$ cm$^{-2}$ is
plotted as a function of their X-ray luminosity for $z<1$ (solid
line), and compared with data from La Franca et al. (2005, solid squares); the
latter have been corrected for the selection effects as described by the authors.
We plot also the expectations of the AGN synthesis
model for the cosmic X-ray background by Gilli et al. (2007, dots).
We also show as a dashed line the obscured fraction as a function of AGN luminosity
when no AGN feedback is considered; note that this has been computed with the same choice of
disk parameters and therefore it is normalized as to yield the same fraction of obscured
AGNs at low luminosities where the feedback is not effective.
\vspace{0.4cm}}

We stress that the inverse correlation between the fraction of obscured objects and luminosity constitutes a 
key {\it test} for AGN feedback as a source of the
luminosity-dependence of AGN obscuration; indeed, when such a feedback
is not included, an opposite correlation is found (as shown in fig. 3) since more luminous AGNs result from a 
larger fraction of cold gas available for both accretion and obscuration.
Note that qualitatively similar results are obtained for a range of AGN feedback efficiencies  
$2\,10^{-2}\leq \epsilon_{AGN}\leq 0.2$. Values outside this range would primarily  violate  
constraints set by galaxy colors, the $M_{BH}-\sigma$ correlation, and the specific star formation 
rate of massive galaxies at high redshifts. 

We expect the overall absorption to evolve strongly
with redshift, as shown in fig. 4 where we plot as a
function of $z$ the fraction of AGN (with any luminosity)
absorbed by gas with column density $N_H\geq 10^{23}$ cm$^{-2}$.
Such a fraction of absorbed AGNs rises appreciably with redshift, in
agreement with the observational findings by La Franca et al.
(2005). Although there is still some disagreement 
among observers on whether the AGN obscuration is purely luminosity dependent 
or both luminosity and redshift dependent (see Gilli et al. 2007 and references therein), 
the latter case constitutes 
a generic expectation of models in which obscuration is generated 
on host galaxy scales. This is due to
the larger fraction of cool galactic gas available for both
BH accretion and AGN absorption at high redshifts; in turn, this
is due to the rapid gas cooling corresponding to the higher
densities of the collapsed structures hosting the AGNs, and to the
frequent galaxy merging events which allow to continuously replenish
the cold gas to be turned into stars at high redshift.
Note that in our model the fraction of
obscured objects with $N_H\geq 10^{22}$ cm$^{-2}$ plotted in fig. 4
includes Compton-thick sources, since it does not resolve the
inner regions of AGNs. These are the regions where the Compton-thick absorbers are
presumably located according to several observational indications, from the local abundance of
Compton-thick sources (favoring large covering factors and consequently
compact sizes, see Risaliti, Maiolino, Salvati 1999),
to the direct limits from Chandra observations of the Circinus galaxy
(see Guainazzi et al. 2001 and references therein) to the rapid oscillation timescale
of the absorber in NGC1365 (Risaliti et al. 2005a).
In the fig. 4, we tentatively estimate the 
contribution of Compton-thin sources (dashed line) to the obscured fraction
by assuming that the difference between the predicted and the observed
density of AGNs shown in fig. 2 is entirely due to Compton-thick sources. 
Finally, in fig. 4 we give our predictions concerning the  
redshift evolution of the absorbed fraction of AGNs with different luminosities, 
which are shown by the thiner lines. Note how for $z\gtrsim 2$
the luminosity dependence of 
the absorbed fractions weakens for faint AGNs with $L_X\lesssim 3\, 10^{44}$ erg/s, while 
it exhibits a sharp transition to low values $\lesssim 0.5$ 
for brighter AGNs. 

\begin{center}
\vspace{0.cm}
\scalebox{0.45}[0.45]{{\includegraphics{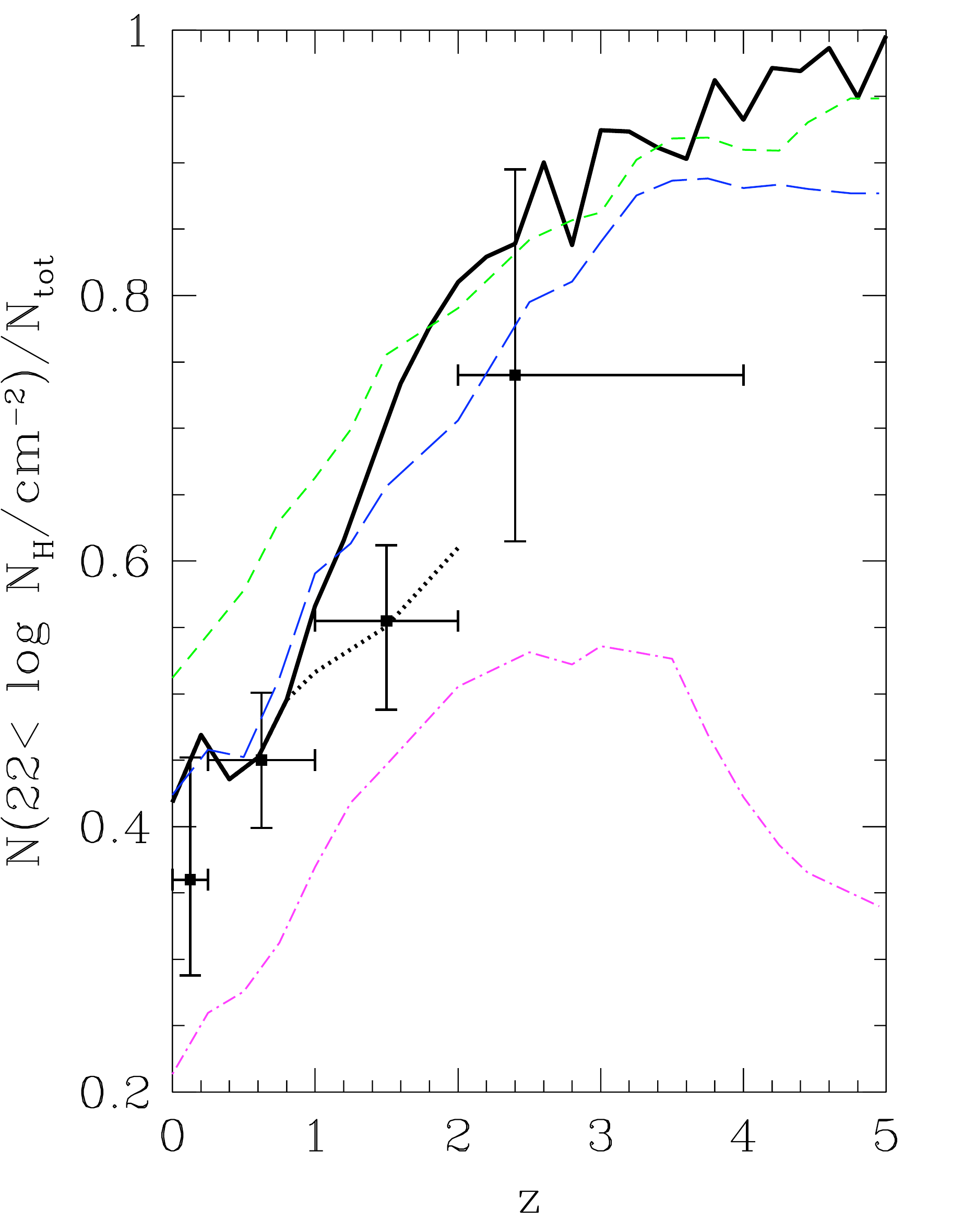}}}
\end{center} {\footnotesize \vspace{-0.1cm }
Fig. 4. - 
The evolution of the predicted
fraction of absorbed ($N_H\geq 10^{22}$ cm$-2$) AGNs for all sources
with  43$\leq$ log $L_X$/erg s$^{-1}\leq$ 46 (solid heavy line). The data from  La Franca
et al. (2005)  have been corrected for the selection effects.
The heavy dotted line represents the contribution to the fraction of absorbed AGNs
due to only Compton-thick sources estimated assuming that the
difference between the observed and the predicted AGN density in fig. 2 is entirely due
to the Compton-thick AGNs. 
The thin solid lines correspond to the absorbed fraction of AGNs within luminosity 
bins of width 0.25 dex centered on the following values: 
$L_X$/erg s$^{-1}=$ 43.25 (green, dashed line), $L_X$/erg s$^{-1}=$ 44.25 (blue, long 
dashed line), $L_X$/erg s$^{-1}=$ 44.7 (cyan, dot dashed line).
\vspace{0.4cm}}

Such a {\it combined} luminosity and redshift dependence of AGN
absorption, which constitutes a {\it specific prediction} of our model,  is 
shown in fig. 5 for a continuous, wide range of luminosities and redshift. 
As redshift increases, the overall fraction of absorbed
AGNs increases, but with a luminosity dependence much
weaker than holding at low $z$ (represented in fig. 2)  
for luminosities $L_X\lesssim 3\, 10^{44}$ erg/s. 
In fact,  the quantity of absorbing gas is predicted to be large enough (due
to the processes discussed above) as to make the expanding
blast wave induced by the AGN feedback ineffective to expell a major
fraction of the galactic gas, which is continuously replenished in
the galactic potential wells. For higher luminosities at $z\gtrsim 2$
a sharp transition occurs between a highly obscured population 
with $L_X\lesssim 3\,10^{44}$ erg/s and a nearly unobscured population 
with $L_X\gtrsim 10^{45}$ erg/s. 
Thus, a testable prediction of our model 
is that for intermediate
luminosities ($10^{43} \lesssim L_X/{\rm erg\,s^{-1}}\lesssim 10^{45}$) 
the ratio of optical to X-ray luminosity functions 
at redshifts $z\gtrsim 2$ should be much {\it lower} than that
at z=0, while for higher luminosities it remains {\it close} to the local value.

\begin{center}
\vspace{0.2cm}
\scalebox{0.8}[0.8]{{\includegraphics{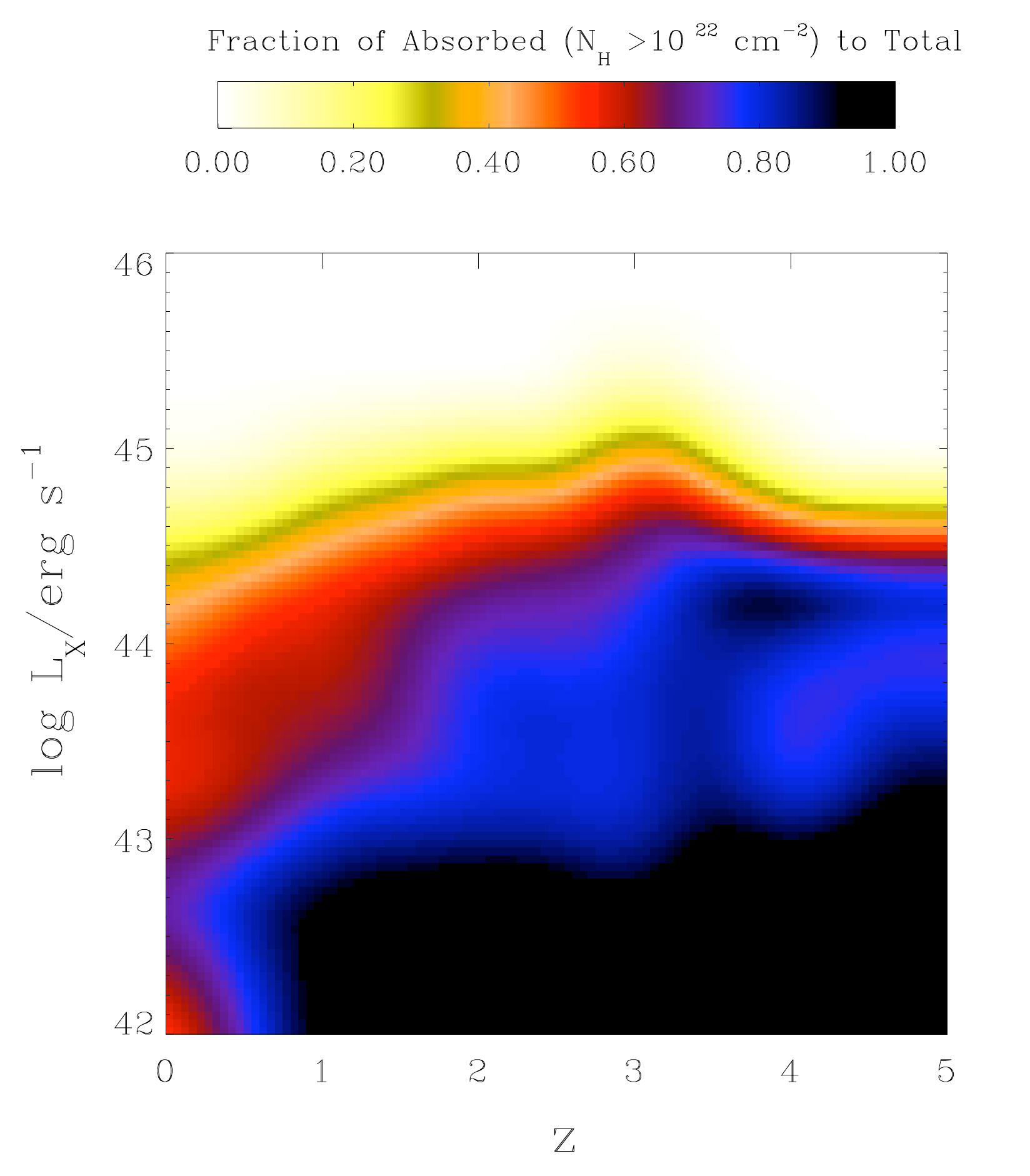}}}
\end{center} {\footnotesize \vspace{0.cm }
Fig. 5. - 
The combined luminosity and redshift dependence of the absorbed 
fraction of AGNs, represented by the color code shown on top.
\vspace{0.4cm}}

\section{Summary and Conclusions}

We have computed the effects of the galactic absorption on AGN emission in a
cosmological context, by including a physical model for AGN fueling and feedback
into a semi-analytic model of galaxy formation in the concordance cosmology.
The model is based on galaxy 
interactions as triggers for AGN accretion and on expanding blast waves as a
mechanism to propagate outwards the AGN energy injected into the interstellar
medium at the center of galaxies.

We have shown that an {\it inverse} dependence of AGN absorption on
luminosity (fig. 3) and a {\it direct} dependence on redshift (fig. 4) is
a natural outcome in such a context. The former arises from the
faster expansion of blast waves induced by feedback of energetic,
luminous AGNs onto the galactic interstellar gas; the rapid sweeping
of gas in the inner regions, where the density was initially
higher, results in the fast formation of a gas-depleted region with 
size larger for higher AGN intrinsic luminosities. Qualitatively 
similar conclusions were reached by Hopkins et al. (2005b) on the basis of
dedicated hydrodynamical N-body simulations.
On the other hand, the redshift dependence of the AGN absorption is due to
the larger amount of cold galactic gas available at high redshift, when the
higher densities allowed for fast cooling occurring in galactic
haloes. The quantitative {\it predictions} of our model are consistent
with existing observations concerning the fraction
of absorbed ($N_H\geq 10^{22}$ cm$^{-2}$) AGNs as a function of their luminosity
and redshift. Our model specifically predicts that 
for AGNs with $L_X\leq 3\,10^{44}$ erg/s
the luminosity dependence of the absorbed fraction  weakens
with increasing redshift (see fig. 5), while 
for the brightest objects with $L_X\gtrsim 3\,10^{44}$
the absorbed fraction quickly decreases with luminosity for $z\gtrsim 2.5$ (see fig. 5). 

Note that, after averaging over the line of sight, unobscured or mildly obscured
AGNs correspond to late stages of the feedback action; in particular, for a given
orientation of the line of sight, the observed column density depends on the time
elapsed since the start of the blast wave expansion. The faster expansion characterizing the blast wave of luminous AGNs thus corresponds to a {\it larger probability} to observe them when the blast has already swept out the central regions of the galaxy IGM.
This picture constitutes an {\it extension} of the unified picture for AGNs (see Antonucci 1993) beyond the canonical scheme based on the single orientation parameter, since the absorption
properties now depend on the combination of {\it orientation} and {\it time} needed to sweep the central regions of the galaxy disk. Note that our picture has also straightforward implications on the connection between star formation and obscuration properties
(proposed , since from our results we on average expect larger star formation in heavily obscured objects, while luminous, mildly absorbed AGNs should be generally associated to galaxies with lesser star formation (or  in transition to a passive state). 
The relevance of evolutionary effects of the kind modeled here in determining the 
absorption properties of AGNs is supported by several observational works. 
Stevens et al.  (2005) and Page et al. (2004) find that X-ray obscured QSOs have 
much higher submillimeter detection rates than X-ray unobscured
QSOs, suggesting strong star formation on-going in the host of  
obscured AGN only.  Sajina et al. (2007) and Martinez-Sansigre et al. (2008) report Spitzer IRS spectra dominated by AGN continuum but showing PAHs features in
emission, typical of starforming galaxies, in samples of ULIRGs and  
radio selected obscured QSOs at z$\sim2$. Lacy et al.  (2007) find evidence for
dust-obscured star formation in type-2 QSOs. From the analysis of the X-ray background, 
Baalntyne ey al. (2006) argue that the AGN obscuration is connected with the star formation in the 
host galaxy. Finally, Martinez-Sansigre et al. (2005, 2008) found little
or no Lyman-$\alpha$ emission in a sample of z$>1.7$ obscured QSOs,
suggesting large scale (kpc) dust distribution. All these findings
are in general agreement with the evolutionary picture, although 
some disagreeing works argue for a dominance of geometrical effect  (see Triester et al. 2008). A 
geometrical effect may be constituted by the gravitational bending of the 
interstellar gas due to the BH, as described in Lamastra et al. (2006); however, except for very massive 
BHs ($M_{BH}\sim 10^9\,M_{\odot}$), this  
will affect mainly regions below our resolution scale of 50 $pc$. Another class of geometrical models, 
based on the luminosity dependence of the sublimation radius of the BH accretion disk,  is that commonly referred to as 
the ''receeding torus'' picture (Lawrence 1991). These
however provide an appreciable luminosity dependence only at very high AGN luminosities $L_x\gtrsim 10^{44-45}$ erg/s 
and for dust located close to the sublimation radius ($1-10$ pc). Dust located in galaxy disk can hardly be affected directly by the AGN radiation. 

In our model we did not try to model the processes leading to
Compton-thick absorption with $N_H\geq 10^{24}$ cm$^{-2}$,
generally thought to be caused by gas directly associated with the
central regions of the AGNs. On the other hand, our model provides a hint
concerning luminosity and redshift dependence to be expected as for the abundance of
Compton-thick sources. Inspection of
fig. 2, where we compare the predicted global (including
Compton-thick sources) density of AGNs with different {\it
intrinsic} luminosities with data corrected for absorption (but not including
Compton-thick sources), shows
that our model predicts a number of low/intermediate-luminosity AGNs
($L_X\leq 10^{44}$ erg/s$^{-2}$) larger than the observed
Compton-thin sources by a factor around 2 at $z\gtrsim 2$. 
While it is possible that our model overestimates the AGN fueling in this 
range of $L_X$ and $z$ (e.g., due to its specific modeling of 
interaction-driven destabilization for cold gas in galactic disks), 
the above excess could support the view that at such luminosities and redshifts 
a fraction around 1/2 of the total AGNs are Compton-thick.
A complementary process which may explain our overprediction  of
low-luminosity AGNs at $z\gtrsim 2$ is suppression of BH growth
in small mass galactic haloes (DM masses $M\leq 10^9\,M_{\odot}$). This may 
be provided by gravitational-rocket effect on the BHs due
to the recoil following the emission of gravitational waves
during the coalescence of BH binaries following galaxy mergers (see Madau \& Quataert 2004).
Such a recoil may produce BH velocities of order $10^2$ km/s,
sufficient to unbind the hole from galaxies with
DM velocity dispersion $\lesssim 50$ km/s.
In this respect, the observational selection of
Compton-thick sources by combining mid-infrared to near-infrared
and optical photometry of galaxies (Fiore et al. 2008a,b) will provide
crucial constrains on the
relative role of obscuration and BH depletion
in low-mass galactic halos.

Our results are also relevant for constraining the physical
mechanisms of AGN feedback onto the interstellar gas, a key issue
in recent developments of cosmological
galaxy formation models. Indeed, the recent realizations of such
models include AGN feedback as the key process to suppress gas cooling
in massive galactic haloes. Actually, two kinds of feedback are at
present implemented in this context: on the one hand, models based
on galaxy interactions as triggers of AGN activity and feedback
relate the latter to the bright, accretion
phase onto supermassive black holes, i.e., to the same phase
corresponding to AGN activity (see Menci et al. 2003, 2007; Di
Matteo et al. 2005; Hopkins et al. 2005a); on the other hand, other
authors associate the AGN feedback only to a quiescent (so called
''radio'') phase of accretion, characterized by very low
accretion rates ($\lesssim 10^{-2}\,M_{\odot}$/yr) and not
observable as radiative AGNs (Bower et al. 2006, Croton et al. 2006). Since
the latter mode would result into a feedback activity continuing down
to low redshift, we expect such models to provide a much milder
dependence of AGN absorption on redshift;  so observational
results on the high-redshift absorption of AGN emission will
constitute an effective test for models of AGN feedback.

Finally, we note that the feedback related to the active AGN phase described
here is effective to decrease the galactic gas and the associated
absorption mainly at low redshift $z\lesssim 2$ and for bright
($L_X\gtrsim 10^{44}$ erg/s) AGNs (see, e.g., Cavaliere \& Menci 2007).
This implies that at high
redshifts the effective cooling and the continuous replenishing of
galactic gas due to fequent merging events will override the depletion
due to the AGN feedback, so the latter can not suppress the
early growth of supermassive BH. Indeed, the predicted number
density of early ($z\geq 4$), bright ($L_x\geq 10^{45.5}$ erg/s)
AGNs shown in fig. 2 is consistent with that observed for bright
($M_i\leq -27.5$) optical QSOs up to $z\approx 6$ (Hopkins et al.
2006). Note however that such a result does not include the BH
spin-up which may occur during the growth due to accretion of gas 
endowed with angular momentum 
(for different modeling of such a process see, e.g., Volonteri
et al. 2005; King, Pringle \& Hofman 2007),
which in turn may yield larger radiative efficiencies
up to values $\eta\approx 0.3$ for a dominant
fraction of BHs; as noted by the above authors, the lower
mass accretion rate related to larger radiation efficiencies may
delay the mass assembly of massive BH at early epochs when included
in a cosmological model.
We shall investigate these issues in a following paper.

\acknowledgments We thank our referee for helpful comments which contributed to improve the paper. We acknowledge grants from ASI. 

\vspace{2cm}
\section*{APPENDIX A}
\setcounter{equation}{0} \renewcommand{\theequation}{A-\arabic{equation}}

The key quantity determining the blastwave evolution is energy deposition ratio $\Delta E/E$  of the 
energy injected up to the time $t$ over to the ISM energy out to the shock radius
at the same time. This is constant in self-similar solutions, which 
imply for the energy injection the law 
$\Delta E(t)\propto t^{2(5-2\omega)/\omega}$ (sect. 5 and Appendix C2 of Lapi, Cavaliere, Menci 2005), normalized to yield 
the total value $\Delta E=\eta\,\epsilon_{AGN}\,\Delta m_{acc}$ (eq. 4) 
at the end of the AGN active phase. Its time behavior is thus related to  the 
exponent $\omega$ defining the decline of the gas density distribution. 

Since the supersonic wave affects the plasma only 
out to the shock distance $R_s(t)$ where the unperturbed (initial) energy content is 
$E(<R_s(t))\propto R_s(t)^{5-2\omega}\propto t ^{2(5-2\omega)/\omega}$ (Lapi et al. 2005), the ratio $\Delta E/E$ is constant, and  constitutes the basic quantity determining the strength of the shock; in other words, 
$\Delta E/E$ remains constant while the wave expands, since both
the energy injected up to the considered time and the ISM energy content within the shock radius grows in time with the 
same law. Thus we choose to compute $\Delta E/E$ as the total  energy injected by the AGN divided into the total
 energy content of the ISM within the whole volume of the disk. 

The luminosity $L$ is the time derivative of $\Delta E$, and thus in the self-similar model 
 is to behave as $L= d \Delta E/dt\propto t^{5(2-\omega)/\omega}$; so its decay is  related to the exponent $\omega$ 
 defining the density decline (see Lapi, Cavaliere, Menci 2005, Sect.5 and Appendix C2). 

For our piecewise approximation of the density profile (point 1 in Sect. 3.2), $L$ is constant during the time when the 
wave sweeps the inner region $r\leq r_d/2$, declines as $L\propto t^{-0.65}$ when the blast has expanded into the region 
$r_d/2< r \leq 3\,r_d/2$, and then drops to zero when $r>3\,r_d/2$, where $\Delta E$ remains constant at the value attained 
at the border $r=3r_d/2$, consistently with the index $\omega=5/2$ holding in that region.

Full consistency with the self-similar blastwave model would require us to assigne each AGN a 
time-dependent luminosity, decaying as described above. However, 
the results are indistingishable from those obtained with the simple prescription $L$=constant 
(provided both luminosities are normalized to the same final total injection $\Delta E$) for two reasons.

1) As for the central $r\leq r_d/2$ region $L$ is indeed constant, 
as we showed above; in the outer region the 
self-similar behavior $L\propto t^{-0.65}$ results in a much slower decay of the luminosity 
compared to the decline of the ISM density encountered by 
the blast $\rho\propto R_s^{-\omega}$. In fact, 
using eq. (5) for the time expansion of $R_s(t)\propto t^{2/\omega}$, 
the density in front of the blast declines as $\rho\propto t^{-2}$. 
Clearly, the regime when the blast expands into the outer region $r>3/2\,r_d$ is not important, since here the luminosity is null and the object is not recognized as an active AGN in computing
the $N_H$ distributions shown in Sect. 4. 

2) Most important, the scale height of the disk $h= r_d/15$ is much shorter than the radius $r_d/2$ enclosing the first radial shell we use to fit the exponential profile of the disk. This in turn means that nearly all 
the random line-of-sights that we draw from the galaxy center only intercept the first radial shell, 
where the luminosity is effectively constant according to the self-similar construction of the model.
Quantitavely, the probability for a line-of-sight to intercepts the outer ($r>r_d/2$) disk 
is $[arctan(2/15)]/(\pi/2)\approx 8\,10^{-2}$. 
Beyond the details of our specific self-similar blastwave model, the physical point is that 
the geometry of the disk is such that the vast majority of the line-of-sights is only affected by 
the inner region; the latter is completely swept out by the blast early on, when the luminosity decay has not yet occurred.

\end{document}